\documentclass[prb,twocolumn]{revtex4}%

\usepackage{graphicx}
\usepackage{amsmath}
\usepackage{ulem}
\usepackage{color}

\newcommand{\be}{\begin{equation}}
\newcommand{\ee}{\end{equation}}
\newcommand{\bea}{\begin{eqnarray*}}
\newcommand{\eea}{\end{eqnarray*}}
\newcommand{\bean}{\begin{eqnarray}}
\newcommand{\eean}{\end{eqnarray}}

\begin{document}

\draft
\title{\bf Topological States in Finite Graphene Nanoribbons Tuned by Electric Fields
}

\author{David M T Kuo}

\address{Department of Electrical Engineering and Department of Physics, National Central
University, Chungli, 32001 Taiwan, China}

\date{\today}

\begin{abstract}
In this comprehensive study, we conduct a theoretical
investigation into the Stark shift of topological states (TSs) in
finite armchair graphene nanoribbons (AGNRs) and heterostructures
under transverse electric fields. Our focus centers on the
multiple end zigzag edge states of AGNRs and the interface states
of $9-7-9$ AGNR heterostructures. For the formal TSs, we observe a
distinctive blue Stark shift in energy levels relative to the
electric field within a range where the energy levels of TSs do
not merge into the energy levels of bulk states. Conversely, for
the latter TSs, we identify an oscillatory Stark shift in energy
levels around the Fermi level. Simultaneously, we reveal the
impact of the Stark effect on the transmission coefficients for
both types of TSs. Notably, we uncover intriguing spectra in the
multiple end zigzag edge states. In the case of finite $9-7-9$
AGNR heterostructures, the spectra of transmission coefficient
reveal that the coupling strength between the topological
interface states can be well controlled by the transverse electric
fields. The outcomes of this research not only contribute to a
deeper understanding of the electronic property in graphene-based
materials but also pave the way for innovations in next-generation
electronic devices and quantum technologies.
\end{abstract}

\maketitle

\section{Introduction}
In the dynamic realm of nanomaterials and nanoelectronics
[\onlinecite{CastroN}-\onlinecite{CahangirovS}], the investigation
of unique electronic properties in graphene-based materials
continues to be a focal point of scientific exploration
[\onlinecite{SonY}-\onlinecite{Kuo3}]. Among these materials,
armchair graphene nanoribbons (AGNRs) have garnered significant
attention. As researchers delve into the manipulation of
electronic states in nanoscale materials, an essential aspect that
demands scrutiny is the Stark shift in AGNRs
[\onlinecite{Novikov}-\onlinecite{NikitaVT}]. Understanding the
Stark shift in AGNRs is not only crucial for tailoring the
electronic properties of these materials but also for unraveling
the intricate interplay between topological phases and Stark
effects [\onlinecite{ZhaoF}-\onlinecite{NikitaVT}].

Although the Stark effects of AGNRs
[\onlinecite{Novikov}-\onlinecite{NikitaVT}] and other
low-dimensional materials [\onlinecite{OkeeffeJ}--
\onlinecite{KuoD}] have been extensively studied, it is poorly
understood for Stark effect on the end zigzag edges of AGNRs
[\onlinecite{ZdetsisAD1}-\onlinecite{KuoDMT}] and the interfaces
of AGNR heterostructures [\onlinecite{CaoT}]. The Stark shift of
these two types of topological states with localized wave
functions offers new avenues for the design and optimization of
nanoelectronic devices. AGNRs are envisioned as key components in
future electronic circuits and transistors
[\onlinecite{BennettPB}]. Moreover, the implications for
optoelectronic applications cannot be
understated[\onlinecite{LinMF}], as the Stark shift in the bound
states of AGNRs may influence absorption and emission properties,
impacting the development of photodetectors and light-emitting
diodes.

This investigation seeks to explore the Stark shift in the end
zigzag edge states and topological interface states of GNRs,
providing insights that could inform future developments in
GNR-based quantum computing
[\onlinecite{ChenCC}-\onlinecite{Bockrath}]. Our study focuses on
a 13-atom-wide AGNR (13-AGNR) and a 9-7-9 AGNR heterostructure, as
illustrated in Figures 1(a) and 1(b), respectively. These
structures were fabricated using the bottom-up synthesis
technique, which is well-suited for producing narrow GNRs
[\onlinecite{LlinasJP}, \onlinecite{DJRizzo}]. Recent studies
[\onlinecite{ZdetsisAD1}-\onlinecite{KuoDMT}] have demonstrated
long 13-AGNR segments with a degeneracy of multiple end zigzag
edge states, distinct from the degeneracy observed in 7-AGNR and
9-AGNR segments. Figures 1(c) and 1(d) show the topological end
zigzag edge states in a long 13-AGNR segment
[\onlinecite{ZdetsisAD1}-\onlinecite{KuoDMT}] and the topological
states arising from the interface states of a 9-7-9 AGNR
heterostructure [\onlinecite{CaoT}]. To investigate the impact of
the Stark shift on the energy levels and transport properties of
these topological states, external electric fields ($F_x$) are
applied along the zigzag edge direction (the x-direction).

We have demonstrated that the energy levels of multiple end zigzag
edge states exhibit nonlinear behavior with respect to the
electric field strength in a short 13-AGNR. Conversely, their
response to electric fields shows linear behavior in a longer
13-AGNR. The spectra of transmission coefficients for various
electric field strengths directly reveal the presence of multiple
zigzag edge states. Meanwhile, the energy levels of topological
interface states in the 9-7-9 AGNR segment exhibit interesting
oscillatory behavior around the Fermi energy. The transmission
coefficient spectra for the 9-7-9 AGNR segment indicate that the
interface states are well-separated from the bulk states and
electrodes. The coupling strength between these interface states
can be finely adjusted by transverse electric fields. Given these
characteristics, the topological interface states of AGNR
heterostructures hold significant promise for the design of
quantum bits [\onlinecite{ChenCC}-\onlinecite{Bockrath}].

\begin{figure}[h]
\centering
\includegraphics[trim=1.cm 0cm 1.cm 0cm,clip,angle=0,scale=0.3]{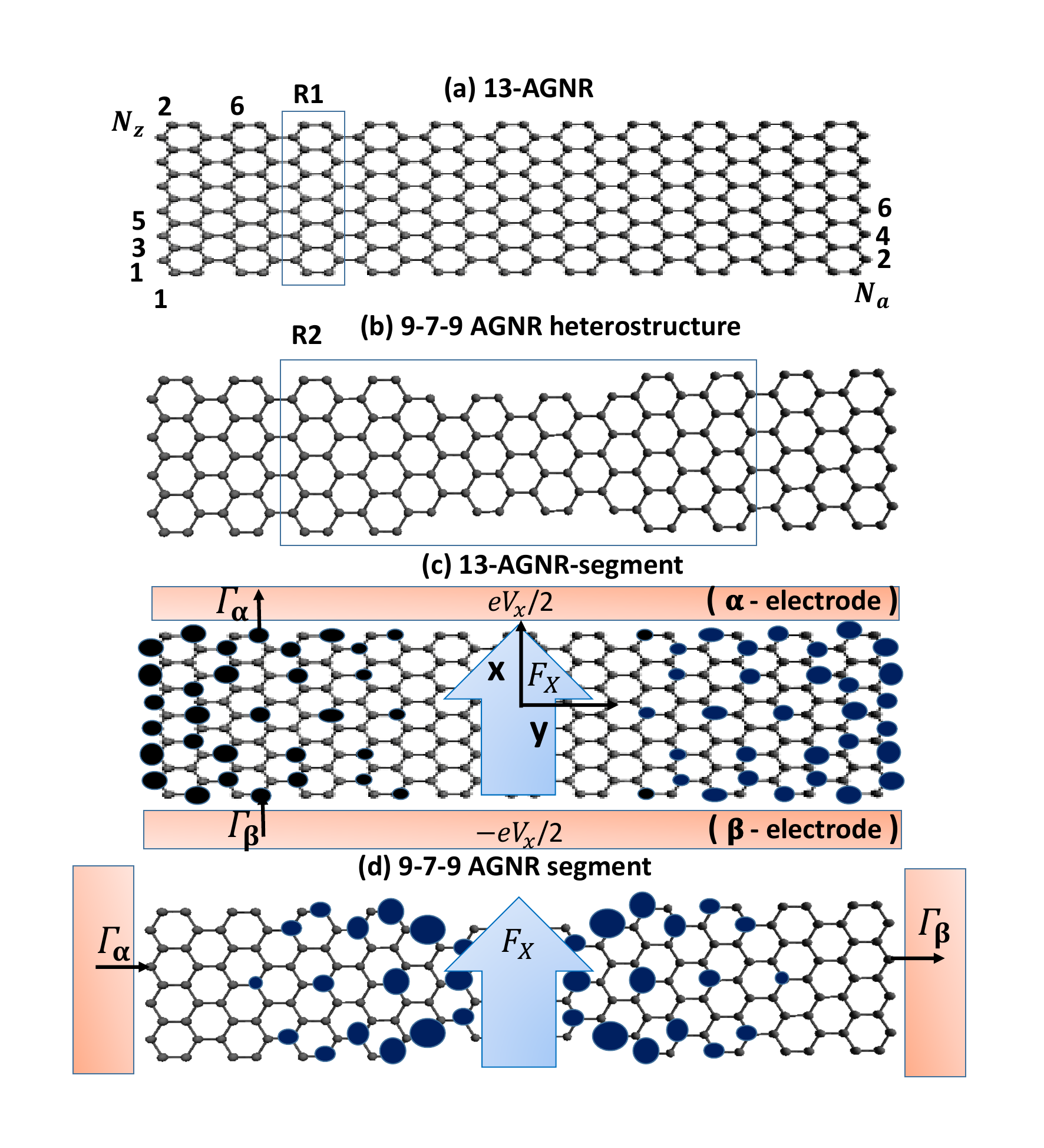}
\caption{(a) Schematic diagram of a 13-atom-wide armchair graphene
nanoribbon (13-AGNR). (b) Structure of a 9-7-9 AGNR
heterostructure. The geometries of the $R_1$ and $R_2$ unit cells
are shown for the 13-AGNR and 9-7-9 AGNR heterostructure,
respectively. (c) A 13-AGNR segment is connected to the $\alpha$
and $\beta$ electrodes. The symbols $\Gamma_{\alpha}$ and
$\Gamma_\beta$ represent the electron tunneling rates between the
$\alpha$ and $\beta$ electrodes and the adjacent atoms at the
armchair edges. (d) A 9-7-9 AGNR heterostructure segment with
zigzag edges coupled to electrodes. The transverse electric field
is given by $F_x = V_x / L_z$, where $L_z =
\sqrt{3}a_{cc}(N_z-1)/2$ and $a_{cc} = 1.42, \text{\AA}$
(carbon-carbon bond length). The x-direction is defined along the
zigzag edge structure of the 13-AGNR and 9-7-9 AGNR segments. The
radius of the circle represents the magnitude of the charge
density for the topological states of the GNRs at $F_x = 0$.}
\end{figure}

\section{Calculation Methodology}
To investigate how external electric fields to influence charge
transport through the finite GNR structure coupled to electrodes,
we employ a combination of the tight-binding model and the Green's
function technique [\onlinecite{Kuo1},\onlinecite{Kuo9}]. The
system Hamiltonian consists of two components: $H = H_0 +
H_{GNR}$. In this context, $H_0$ signifies the Hamiltonian of the
electrodes, which encompasses the interaction between the
electrodes and the GNR. Meanwhile, $H_{GNR}$ represents the
Hamiltonian of GNR, which is expressed as follows:

\begin{small}
\begin{eqnarray}
H_{GNR}&= &\sum_{\ell,j} E_{\ell,j} d^{\dagger}_{\ell,j}d_{\ell,j}\\
\nonumber&-& \sum_{\ell,j}\sum_{\ell',j'} t_{(\ell,j),(\ell', j')}
d^{\dagger}_{\ell,j} d_{\ell',j'} + h.c,
\end{eqnarray}
\end{small}

Here, $E_{\ell,j}$ represents the on-site energy of the orbital in
the ${\ell}$-th row and $j$-th column. The operators
$d^{\dagger}_{\ell,j}$ and $d_{\ell,j}$ create and annihilate an
electron at the atom site denoted by ($\ell$,$j$). The parameter
$t_{(\ell,j),(\ell', j')}$ characterizes the electron hopping
energy from site ($\ell'$,$j'$) to site ($\ell$,$j$). We assign
the tight-binding parameters for GNR as follows:
$E_{\ell,j}=E_0+(\ell-(N_z+1)/2)eV_x/(N_z-1)$, where the applied
voltage $V_x$ responses to the external electric fields
$F_x$[\onlinecite{Pizzochero}]. Meanwhile, we set
$t_{(\ell,j),(\ell',j')} = t_{pp\pi} = 2.7$ eV for the
nearest-neighbor hopping strength. The electron hopping strengths
for second- and third-neighbor interactions can be considered as a
perturbative effect [\onlinecite{LopezSancho}]. The calculated
energy levels of the topological states, under this approximation,
are in good agreement with results obtained from first-principles
method (DFT) calculation [\onlinecite{DJRizzo},
\onlinecite{Zdetsis}]. For simplicity, this study neglects these
perturbative terms as well as electron Coulomb interactions within
the topological states. However, the latter effect may induce a
magnetic order parameter for the topological states in the chain
[\onlinecite{CaoT}].


The bias-dependent transmission coefficient (${\cal
T}_{\alpha,\beta}(\varepsilon,V_x)$) of a GNR connected to
electrodes can be calculated using the formula ${\cal
T}_{\alpha,\beta}(\varepsilon) =
4Tr[\Gamma_{\alpha}(\varepsilon)G^{r}(\varepsilon)\Gamma_{\beta}(\varepsilon)G^{a}(\varepsilon)]$
[\onlinecite{Kuo2},\onlinecite{Kuo3}], where
$\Gamma_{\alpha}(\varepsilon)$ and $\Gamma_{\beta}(\varepsilon)$
denote the tunneling rate (in energy units) at the $\alpha$ and
$\beta$ leads, respectively, and ${G}^{r}(\varepsilon)$ and
${G}^{a}(\varepsilon)$ are the retarded and advanced Green's
functions of the GNRs coupled to the electrodes, respectively. The
tunneling rates are determined by the imaginary part of the
self-energy originating from the coupling between the $\alpha$
($\beta$) electrode and its adjacent GNR atoms. In terms of
tight-binding orbitals, $\Gamma_{\alpha}(\varepsilon)$ and Green's
functions are matrices. For simplicity,
$\Gamma_{\alpha(\beta)}(\varepsilon)$ for interface atoms between
the electrodes and the GNR possesses diagonal entries with a
common value of $\Gamma_t$ [\onlinecite{Kuo1},\onlinecite{Kuo9}].

\section{Results and discussion}
\subsection{Electronic Structures of AGNRs with Transverse Electric Fields}
In Figures 2(a)-2(d), we present the electronic structures of
13-AGNRs shown in Fig. 1(a) for various $V_x$ values. As depicted
in Fig. 2, the application of a transverse electric field induces
a transition in AGNRs from a semiconductor to a semimetal state.
Several intriguing phenomena arise upon the introduction of $V_x$:

(a) The band gap of AGNRs ($E_g = 0.714$ eV) diminishes with
increasing $V_x$ values. Notably, when $V_x = 5.4$ V, two
Fermi-Dirac coves near the Fermi level ($\varepsilon_F$) are
observed in Fig. 2(d). This modulation of electronic structures in
AGNRs by an electric field has been documented by various research
groups employing different methodologies
[\onlinecite{Hassan}-\onlinecite{Pizzochero}]. Furthermore, the
semiconductor-to-semimetal phase transition illustrated in Fig. 2
is also applicable to AGNRs with different widths
[\onlinecite{Pizzochero}]. The reduction in the band gap as a
function of increasing electric field for over 30 AGNRs, with
widths ranging from $N_z = 11$ to $N_z = 43$, was theoretically
reported in reference [\onlinecite{Pizzochero}].

(b) The second derivative of the curves for the first conduction
subband indicates that electron effective masses increase with the
strength of the transverse electric field in the small field
regions [\onlinecite{Hassan}].

(c) The flat bands at $\varepsilon -\varepsilon_F = \pm 2.7$ eV
for $V_x = 0$ undergo significant alterations for large $V_x \ge
1.8$ V. Analytical proofs have demonstrated the existence of flat
bands in AGNRs for $N_z$ with odd numbers [\onlinecite{ZhengHX}].

\begin{figure}[h]
\centering
\includegraphics[angle=0,scale=0.3]{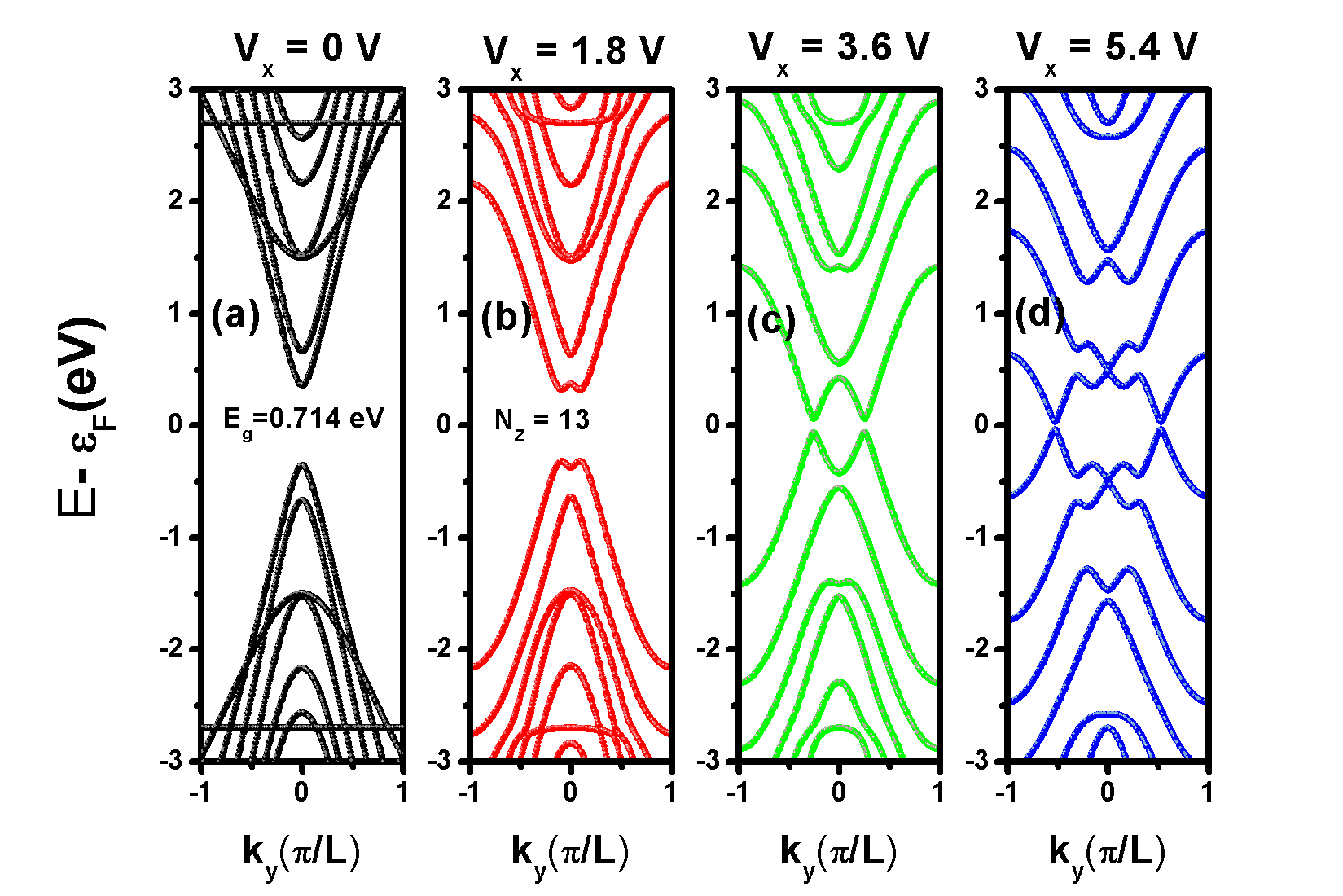}
\caption{Electronic subband structures for various $V_x$ values in
13-AGNR: (a) $V_x = 0$, (b) $V_x = 1.8$ V, (c) $V_x = 3.6$ V, and
(d) $V_x = 5.4 $ V. Here $L = 1$ u.c.}
\end{figure}

\subsection{Finite AGNRs with Transverse Electric Fields}
In Fig. 2, the electronic states of AGNRs demonstrate the Stark
effect on the bulk states. To elucidate the Stark effect on the
end zigzag edge states of AGNRs, we must consider finite AGNRs. We
present the calculated energy levels of finite 13-AGNRs as
functions of applied voltage ($V_x$) for various lengths of AGNRs
($N_a$) in Fig. 3. For simplicity, we set $\varepsilon_F = 0$
throughout this article. In the absence of electric fields, four
energy levels labeled as $\Sigma_{c,1}$, $\Sigma_{c}$,
$\Sigma_{v}$, and $\Sigma_{v,1}$ exist within the energy range of
$|E(eV)|\le 0.5$ eV in Fig. 3(a). The energy levels $\Sigma_{c,1}$
and $\Sigma_{c}$ ($\Sigma_{v}$ and $\Sigma_{v,1}$) almost merge
for $N_a = 84$ ($L_a = 8.8$ nm), with a difference of
$\Sigma_{c,1}-\Sigma_c = 1.72$ meV. Such behavior was also
observed in previous work [\onlinecite{ZdetsisAD}], where the
author calculated the energy levels of 13-AGNRs using the DFT. In
reference [\onlinecite{ZdetsisAD}], the degeneracy of
$\Sigma_{c,1}$ ($\Sigma_{v,1}$) and $\Sigma_c$ ($\Sigma_{v}$) is
referred to as multiple end zigzag edge states for large $N_a$
values. As shown in Fig. 3(c) and 3(d), the bulk states exhibit
Franz-Keldysh behavior with respect to the electric field when
$V_x \ge 1.8$ V. In Fig. 3(d), it is difficult to distinguish the
curve of $\Sigma_{c,1}(\Sigma_{v,1})$ from
$\Sigma_{c}(\Sigma_{v})$. These curves show a linear function of
$V_x$ before merging into the bulk states.

\begin{figure}[h]
\centering
\includegraphics[angle=0,scale=0.3]{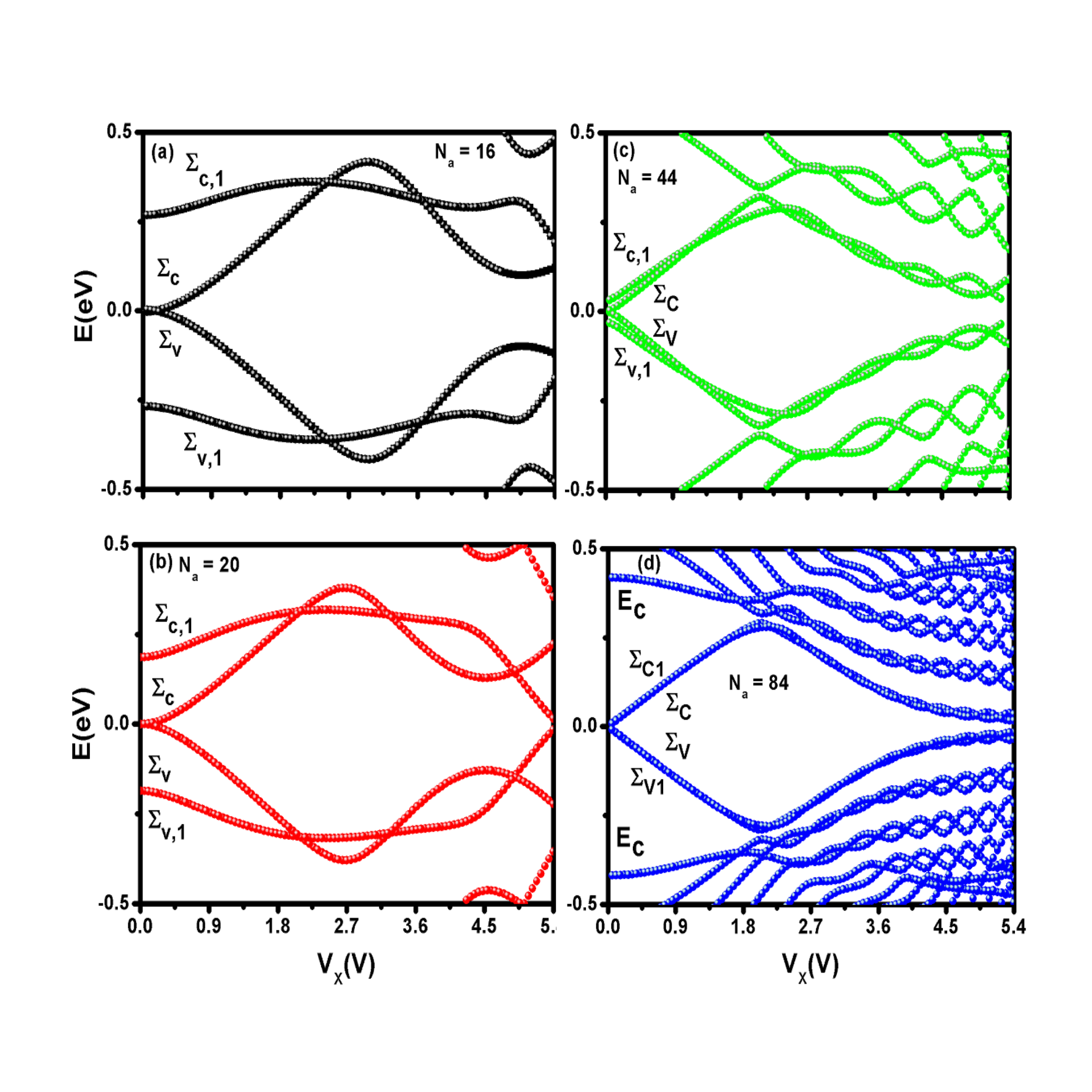}
\caption{Energy levels of finite 13-AGNRs as functions of applied
voltage $V_x$ for various $N_a$ values. (a) $N_a = 16 $ ($L_a =
1.56$ nm), (b) $N_a = 20 $ ($L_a = 1.99$ nm), (c) $N_a = 44 $
($L_a = 4.54$ nm), and (d) $N_a = 84$ ($L_a = 8.8$ nm).}
\end{figure}

To further clarify the Stark effect on the energy levels of end
zigzag edge states shown in Fig. 3, we present the calculated
charge density distribution $|\psi(\ell,j)|^2$ of finite 13-AGNRs
with $N_a = 44$ in Fig. 4. In Fig. 4(a), the charge density
($|\psi(\ell,j)|^2$) of $\Sigma_c$ shows exponential decay in the
armchair directions, resulting in its charge densities being most
distributed at the end zigzag edge sites. Maximum charge densities
appear at sites ($\ell = 6, j = 1(j=44)$) and ($\ell = 8,
j=1(j=44)$). As depicted in Fig. 4(b), the charge density
$|\psi(\ell,j)|^2$ of $\Sigma_{c,1}$ exhibits maximum values at
sites ($\ell = 4, j = 1(j=44)$) and ($\ell = 10, j = 1(j=44)$).
The wave function of $\Sigma_{c,1}$ along the y-direction decays
slower than that of $\Sigma_c$. In the presence of $F_x$, the
charge densities of $\Sigma_c$ ($\Sigma_v$) are pushed to higher
(lower) potential regions in Fig. 4(c) (Fig. 4(d)), indicating
that the applied electric field aligns with the dipole moment of
electronic states ($\Sigma_c$ and $\Sigma_v$) [\onlinecite{LiDH}].
This explains the blue Stark shift of the end zigzag edge states
(shown in Fig. 3(c) and 3(d)).

\begin{figure}[h]
\centering
\includegraphics[angle=0,scale=0.3]{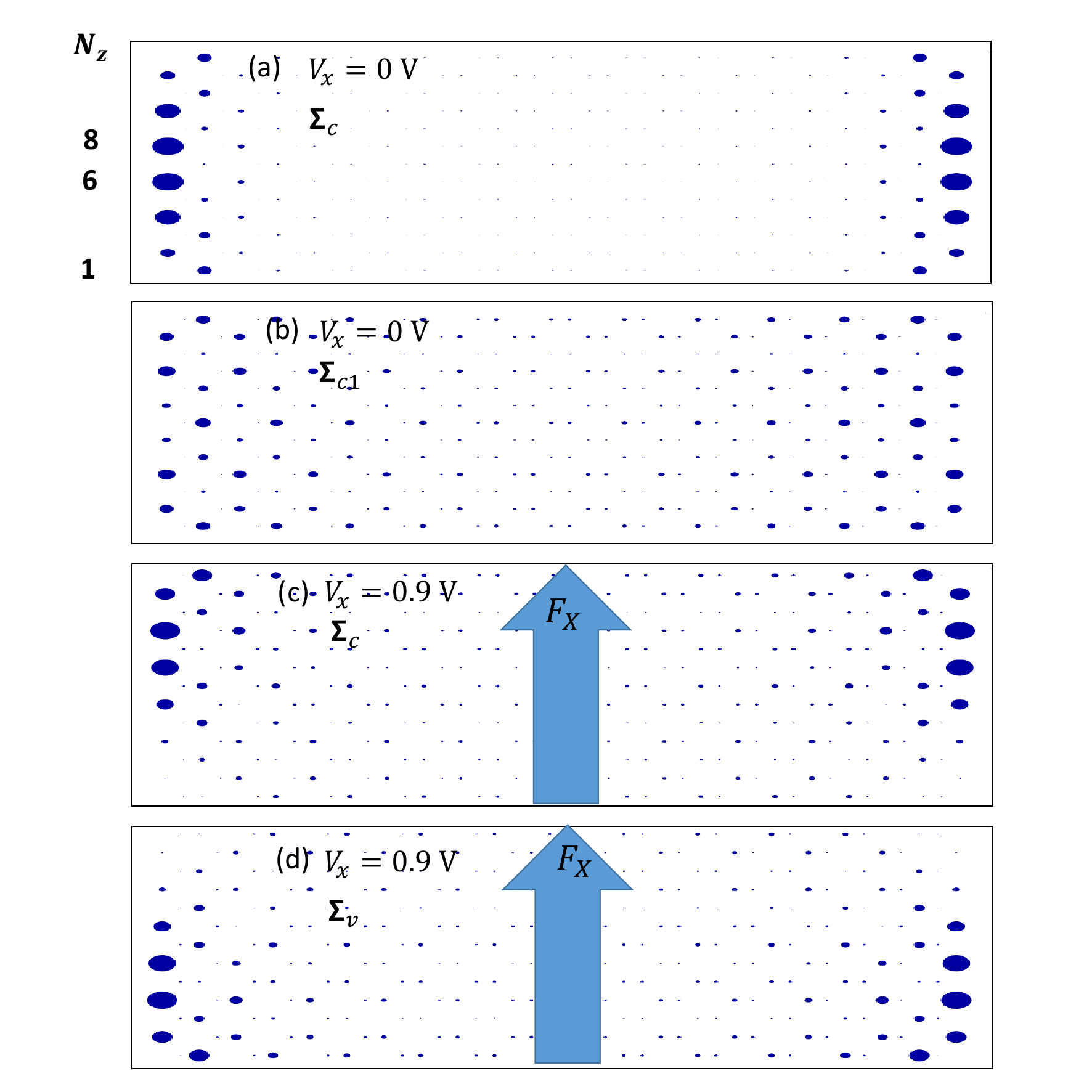}
\caption{Charge density ($|\Psi(\ell,j)|^2$) of 13-AGNRs with $N_a
= 44 $. Diagrams (a)-(d) represent the charge densities of
$\Sigma_c$ at $V_x = 0$, $\Sigma_{c,1}$ at $V_x =0$, $\Sigma_c =
0.141$ eV at $V_x = 0.9$ V, and $\Sigma_v = -0.141$ eV at $V_x =
0.9$ V, respectively.}
\end{figure}

To provide more direct evidence of multiple end zigzag edge
states[\onlinecite{ZdetsisAD}], we present the calculated
transmission coefficient ${\cal T}_{\alpha,\beta}(\varepsilon)$ of
finite 13-AGNRs with $N_a = 84$ and $\Gamma_t = 0.09$ eV in Fig.
5(a)-5(f), where we consider the armchair edges of AGNRs
interconnected to the electrodes as shown in Fig. 1(c). As
depicted in Fig. 5(a), the quantum confinement effect of finite
AGNRs enlarges the gap $E_c - E_v = 0.836$ eV, which is larger
than $E_g = 0.714$ eV in infinitely long 13-AGNRs (see Fig. 2(a)).
Additionally, there exists a significant peak ($\Sigma_0$) in the
middle gap for $V_x = 0$. Although there are four energy levels
around $\varepsilon = 0$ [\onlinecite{Zdetsis}], notably the
maximum ${\cal T}_{\alpha,\beta}(\varepsilon = 0)$ is smaller than
two. To resolve these four energy levels, one must consider very
weak coupling strength between the AGNRs and the electrodes (e.g.,
$\Gamma_t = 9$ meV). In Fig. 5(b), $\Sigma_0$ splits into
$\varepsilon_{C}$ and $\varepsilon_{V}$. The peaks labeled by
$\varepsilon_C$ and $\varepsilon_V$ are larger than one. Such
magnitudes of peaks larger than one are enhanced with increasing
electric fields, as seen in Fig. 5(c)-5(f). The enhancement of
$\varepsilon_{C(V)}$ larger than one is evidence of multiple end
zigzag edge states of 13-AGNRs.

\begin{figure}[h]
\centering
\includegraphics[angle=0,scale=0.3]{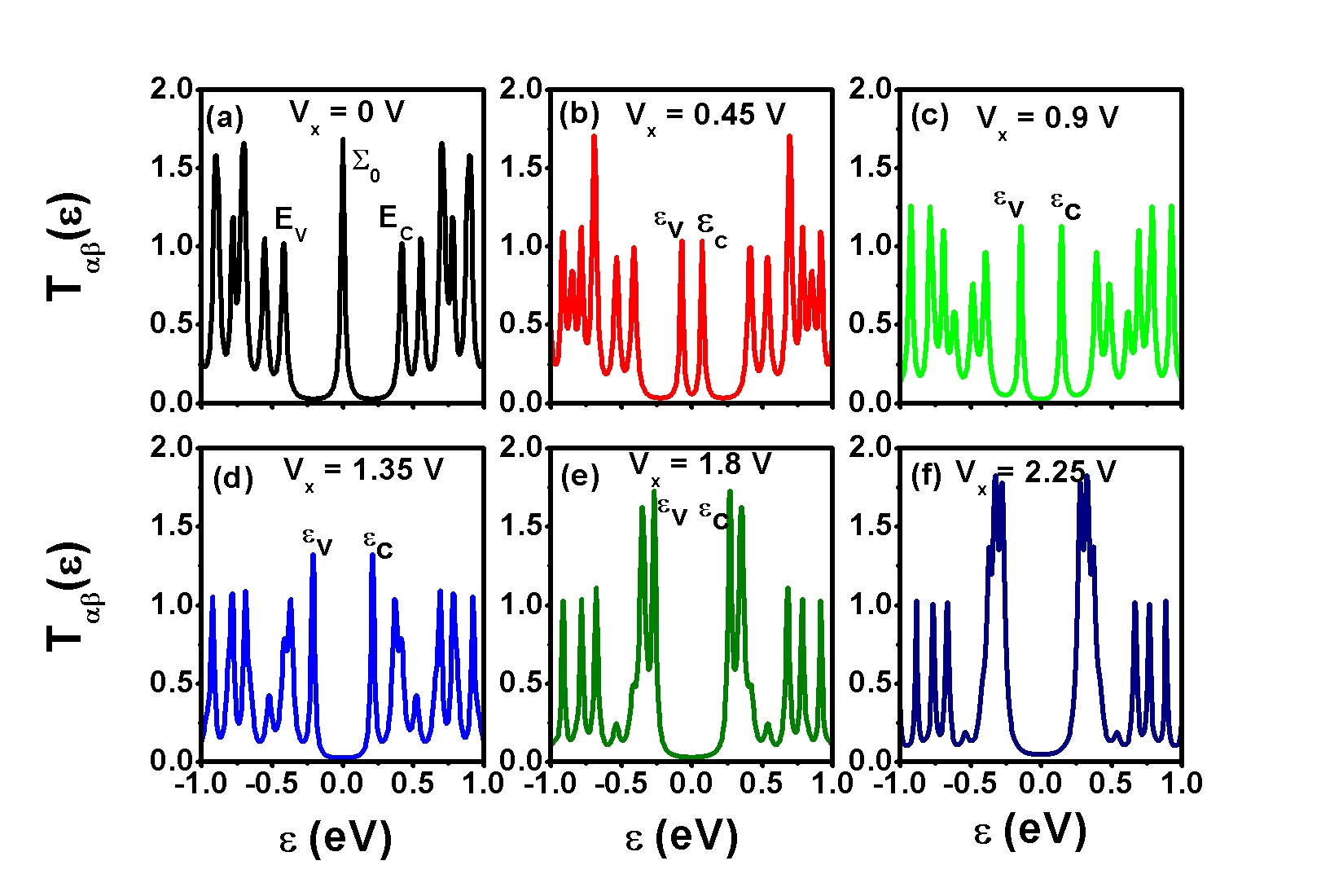}
\caption{ Transmission coefficient ${\cal
T}_{\alpha,\beta}(\varepsilon)$ of 13-AGNR with $N_a = 84$ and
$\Gamma_t = 90$ meV for various $V_x$ values. (a) $V_x = 0$, (b)
$V_x = 0.45$ V, (c) $V_x = 0.9$ V, (d) $V_x = 1.35$ V, (e) $V_x =
1.8$ V, and (f) $V_x = 2.25$ V.}
\end{figure}

\subsection{Electronic Structures of $9_2-7_3-9_2$ AGNR Superlattice with Transverse Electric Fields}
To investigate the Stark effect on the topological states (TSs) at
the interfaces between the 9-AGNR and 7-AGNR segments, we present
the calculated electronic structures of $9_{w}-7_{x}-9_{w}$ AGNR
superlattices (SLs) shown in Fig. 1(b) for various transverse
electric fields in Fig. 6. The notations $w = 2$ and $x = 3$
denote the lengths of 9-AGNR and 7-AGNR, respectively, in units of
unit cells (u.c) of AGNRs. In Fig. 6(a), we observe subbands
around the Fermi energy [\onlinecite{CaoT}], which have also been
confirmed experimentally in a recent report
[\onlinecite{DJRizzo}]. The topological subband structures without
$V_x$ can be well described by the Su-Schrieffer-Heeger (SSH)
model [\onlinecite{SSH},\onlinecite{ObanaD}], which provides an
analytical expression of $E(k) = \pm
\sqrt{t^2_x+t^2_w-2t_xt_w~cos(k(\pi/L))}$. Here, $t_x$ and $t_w$
represent the electron hopping strengths in the 7-AGNR segment and
9-AGNR segment, respectively, while $L = 7 $ u.c denotes the
length of the super unit cell [\onlinecite{Kuo3}].

In Fig. 6(a), the gap of the minibands formed by TSs is $\Delta =
76.8$ meV. This band gap exhibits oscillatory behavior with
respect to $V_x$, as depicted in Fig. 6. Specifically, we have
$\Delta = 65$ meV, $\Delta = 31.7$ meV, $\Delta = 31.7$ meV,
$\Delta = 7.26$ meV, and $\Delta = 88.5$ meV for $V_x = 0.9$ V,
$V_x = 1.8$ V, $V_x = 2.7$ V, $V_x = 3.6$ V, and $V_x = 4.5$ V,
respectively. Even at $V_x = 4.5$ V, the topological subbands
remain well separated from the bulk subbands. This unique
characteristic suggests that the interface TSs can avoid noise
from the bulk subbands, such as electron inter-subband transitions
under optical phonon-assisted processes.

\begin{figure}[h]
\centering
\includegraphics[angle=0,scale=0.3]{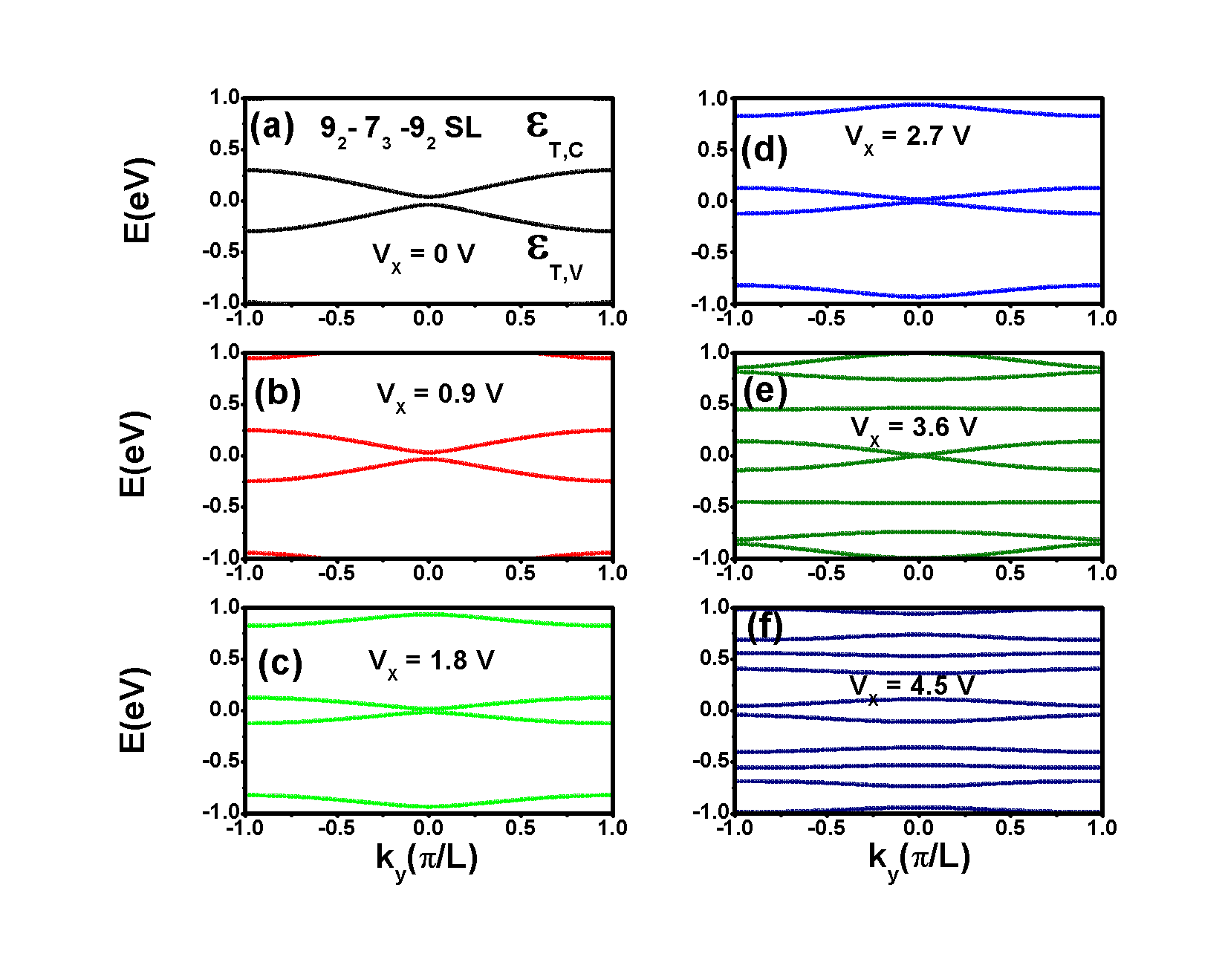}
\caption{Electronic structure of $9_2-7_3-9_2$ AGNR superlattice
for various voltages. Diagrams (a)-(f) correspond to $V_x = 0$ V,
$V_x = 0.9$ V, $V_x = 1.8$ V, $V_x = 2.7 $ V, $V_x = 3.6 $ V, and
$V_x = 4.5 $ V. Here $L = 7$ u.c.}
\end{figure}

\subsection{ $9_w-7_x-9_w$ AGNR Segments with Transverse Electric Fields}
Although the 9-7-9 AGNR SLs fabricated using the bottom-up
synthesis technique are limited in length by the 7-AGNR segment
with $x = 3$ [\onlinecite{DJRizzo}], we focus on the topological
interface states of $9_w$-$7_x$-$9_w$ AGNR segments with $x = 3$
and $x = 5$ in the following discussion. In a finite $9_w-7_3-9_w$
AGNR segment, the interface states represent robust topological
states (TSs) resistant to defect scattering and other excitation
modes due to their localized nature[\onlinecite{DJRizzo}]. To
elucidate the Stark shift of these TSs, we present the calculated
energy levels of $9_w-7_3-9_w$ AGNR segments as functions of $V_x$
for two different lengths in Fig. 7. We consider a $9_4-7_3-9_4$
AGNR segment with $N_a = 44$ in Fig. 7(a). Four energy levels
labeled by $\Sigma_c$, $\Sigma_v$, $\varepsilon_{T,C}$, and
$\varepsilon_{T,V}$ are observed in the absence of an electric
field. Here, $\Sigma_c = 0.18$ meV and $\Sigma_v = -0.18$ meV
arise from the two end zigzag edge states, while
$\varepsilon_{T,C} = 0.1167$ eV and $\varepsilon_{T,V} = -0.1167$
eV arise from the two interface states between the 9-AGNR and
7-AGNR segments [\onlinecite{Kuo3}]. These two types of TSs
exhibit oscillatory Stark shift phenomena with respect to $V_x$.
Such behaviors distinguish from the blue Stark shift of multiple
end zigzag edge states of 13-AGNR shown in Fig. 3(c). To
understand the Stark shift of AGNRs with different $N_z$ from $N_z
= 13$, we also calculate the end zigzag edge states of AGNR with
$N_z = 9$ and $N_a = 44$ as functions of $V_x$ (blue curve) in
Fig. 7(a). This blue curve also exhibits oscillatory Stark shift.

In Fig. 7(b), we present the Stark shift of $9_8-7_3-9_8$ AGNR
segment. Because the 7-AGNR segment remains as 3 unit cells, the
energy levels of $\varepsilon_{T,C}$ and $\varepsilon_{T,V}$ are
unchanged at $V_x = 0$. Meanwhile, $\varepsilon_{T,C}$ and
$\varepsilon_{T,V}$ show a red Stark shift for $V_x$ smaller than
2 V. To illustrate the red Stark shift phenomena, we present the
charge densities of $\varepsilon_{T,C}$ and $\varepsilon_{T,V}$ at
$V_x = 0.918$ V in Fig. 8. The charge density of
$\varepsilon_{T,C}$ accumulates in the lower potential area, while
that of $\varepsilon_{T,V}$ accumulates in the higher potential
area. The electric field opposes the dipole moment of electronic
states, explaining their red Stark shift at $V_x = 0.918$ V. The
charge densities of $\varepsilon_{T,C}$ and $\varepsilon_{T,V}$
mainly distribute at the interfaces between 9-AGNR and 7-AGNR
segments. The mixing between the end zigzag edge states and the
interface states is very small when the 9-AGNR segments are
enlarged. Note that in the scenario of a $9_4-7_3-9_4$ AGNR
segment, the mixing between the end zigzag edge states and the
interface states can be large in the presence of electric fields.

\begin{figure}[h]
\centering
\includegraphics[angle=0,scale=0.3]{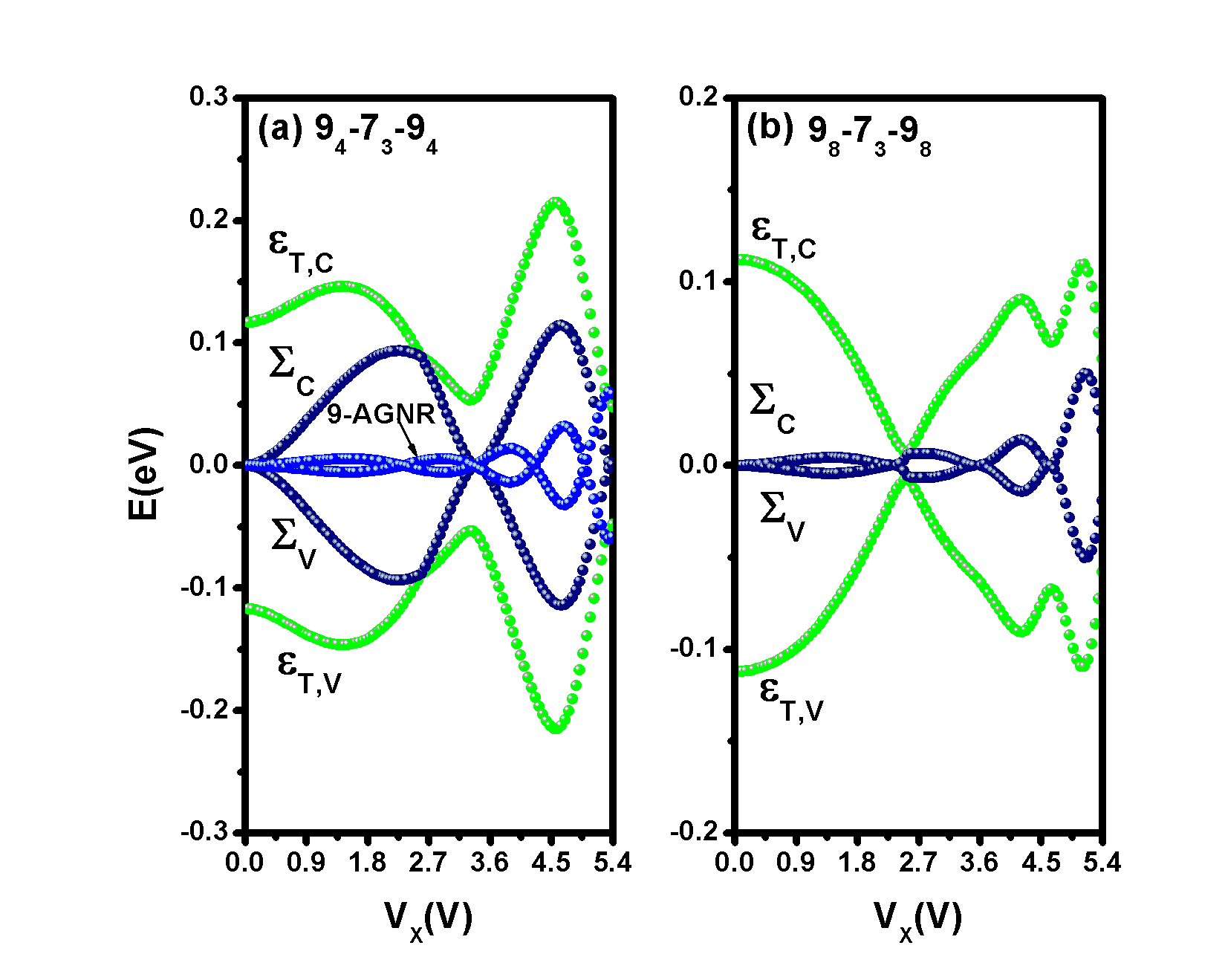}
\caption{Energy levels of 9-7-9 AGNR heterostructures as functions
of $V_x$ for (a) $9_4-7_3-9_4$  and (b) $9_8-7_3-9_8.$}
\end{figure}

\begin{figure}[h]
\centering
\includegraphics[angle=0,scale=0.25]{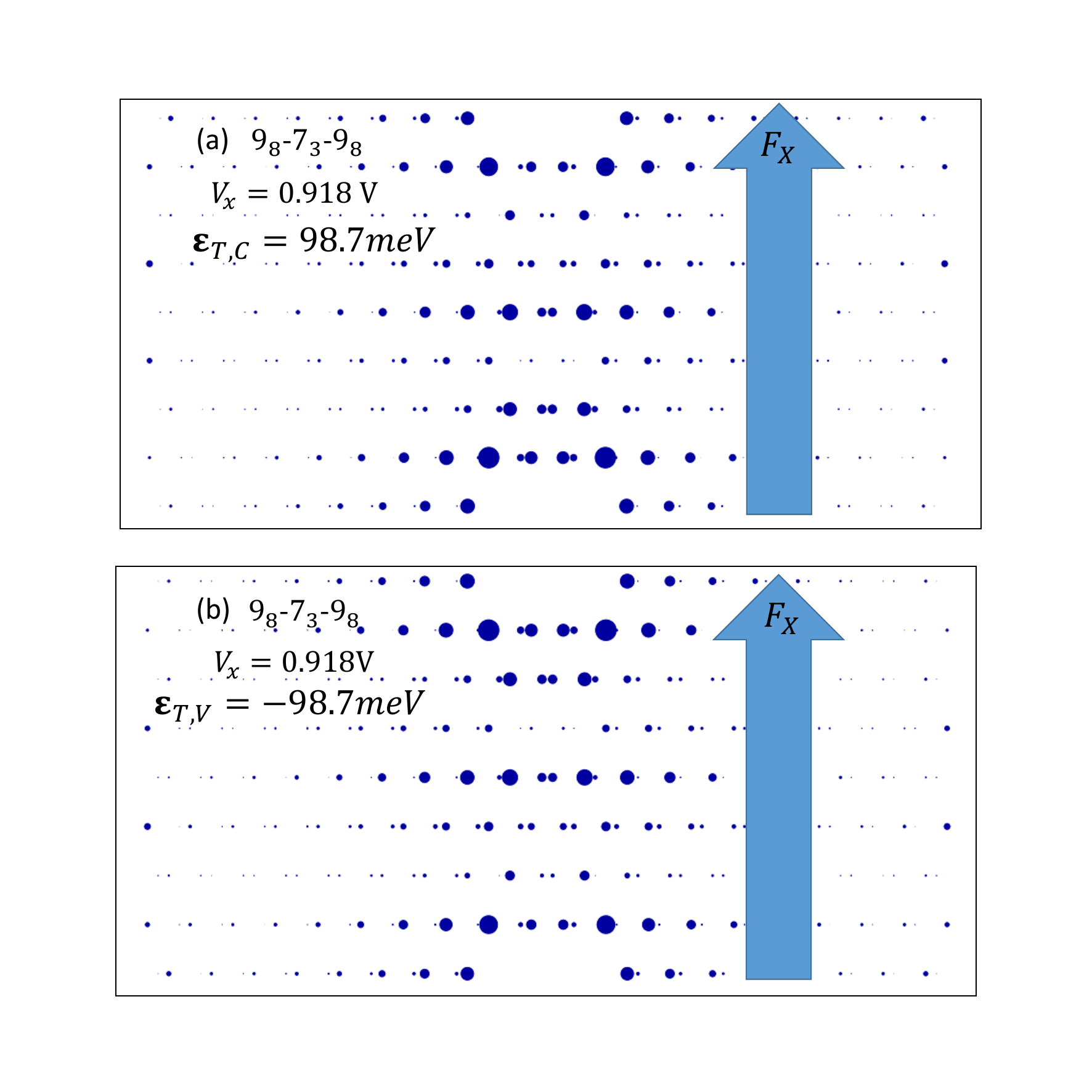}
\caption{Charge density ($|\psi(\ell,j)|^2$) of $9_8-7_3-9_8$ AGNR
segment. (a) and (b) denote, respectively, the charge densities of
$\varepsilon_{T,C} = 98.7$ meV and $\varepsilon_{T,V} = -98.7$~meV
at $V_x = 0.918$ V.}
\end{figure}

In Figs. 7(a) and (b), the length of the 7-AGNR segment remains 3
unit cells. To clarify the length effect of the 7-AGNR segment, we
present the energy levels of a $9_8-7_5-9_8$ AGNR heterostructure
as functions of $V_x$ in Fig. 9. Because the length of the 7-AGNR
segment is 5 unit cells, the value of $t_x$ is reduced to $-38$
meV. The energy levels of $\varepsilon_{T,C(V)}$ and
$\Sigma_{C(V)}$ continue to exhibit oscillatory behavior around
the Fermi energy. We conclude that the oscillatory Stark shift of
$\varepsilon_{T,C(V)}$ and $\Sigma_{C(V)}$ are essential
characteristics of the TSs of $9-7-9$ AGNR heterostructures.

\begin{figure}[h]
\centering
\includegraphics[angle=0,scale=0.3]{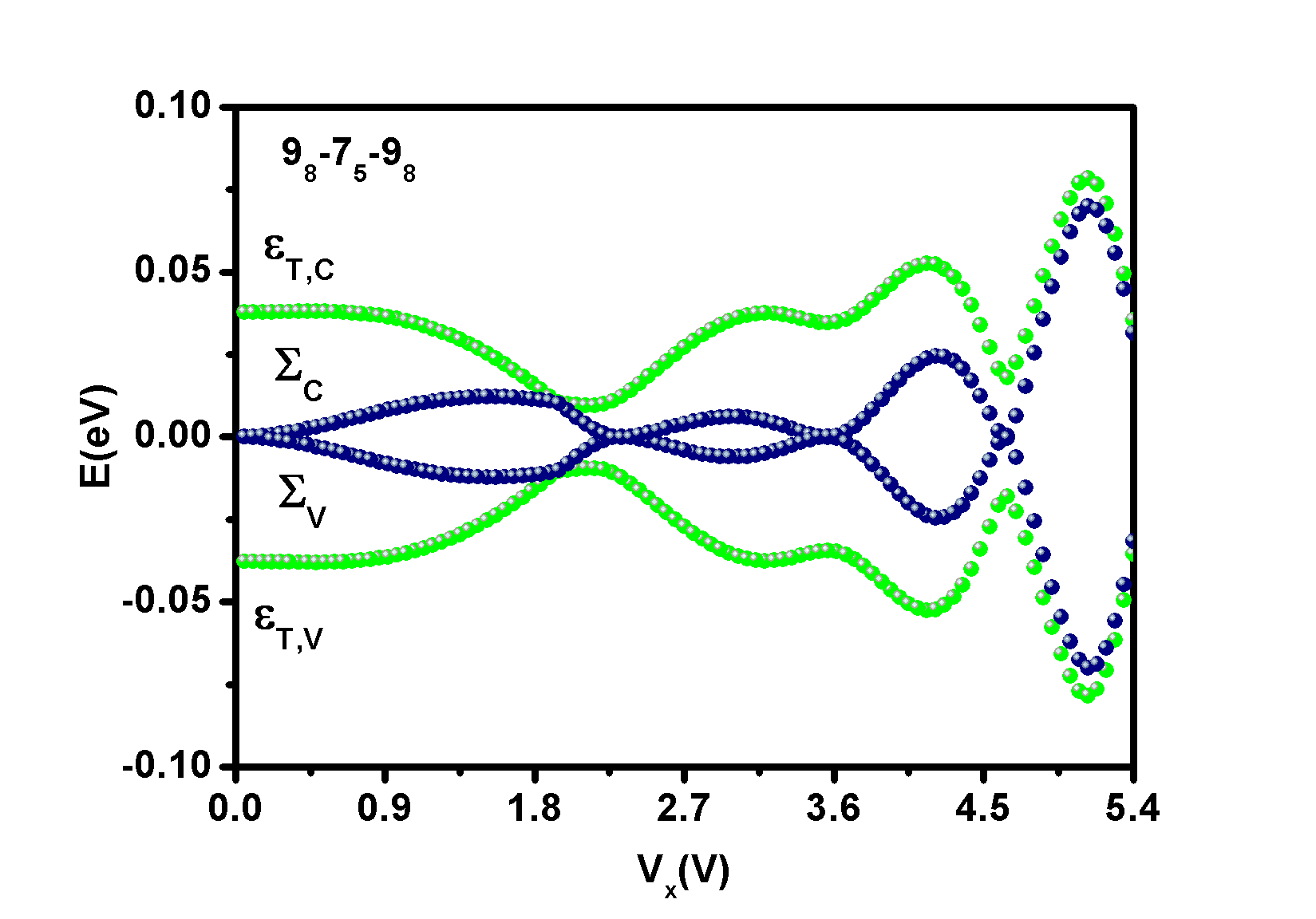}
\caption{Energy levels of $9_8-7_5-9_8$ AGNR segment with $L_a =
8.8$ nm as functions of $V_x$.}
\end{figure}

Next, we present the transmission coefficient ${\cal
T}_{LR}(\varepsilon)$ of $9_8-7_5-9_8$ AGNR segment with zigzag
edges coupled to the electrodes as shown in Fig. 1(d). Fig. 10(a)
shows the transmission coefficient for the case of $V_x = 0$. Note
that in the calculation of Fig. 10, we have considered $\Gamma_t =
0.81$ eV to reveal the charge transport through the interface TSs.
Because the amplitudes of charge densities for $\varepsilon_{T,C}$
and $\varepsilon_{T,V}$ are small near the end zigzag edge sites,
the effective tunneling rates between the electrodes and the
interface TSs are small [\onlinecite{Kuo2}]. This implies that it
is difficult to reveal the spectra of $\varepsilon_{T,C}$ and
$\varepsilon_{T,V}$ for small $\Gamma_t$ values. Although there
exist the end zigzag edge states for finite $9_8-7_5-9_8$ AGNR
heterostructures, charges cannot transport through such localized
states for $\Gamma_t = 0.81$ eV. As depicted in Figs. 10(b)-10(f),
the energy level separation $\varepsilon_{T,C}-\varepsilon_{T,V} =
\Delta = |2t_x|$ reveals the oscillatory Stark shift phenomena of
Fig. 9. The results of Fig. 10 indicate that the energy levels of
TSs resulting from the interface states between the 9-AGNR and the
7-AGNR segments can be well modulated by the transverse electric
fields. Based on the results shown in Fig. 10, the topological
interface states of $9_w$-$7_x$-$9_w$ AGNR segments effectively
function as a series of double quantum dots. Therefore, they can
be used to implement charge qubits designed with graphene quantum
dots [\onlinecite{ChenCC}-\onlinecite{Bockrath}].

\begin{figure}[h]
\centering
\includegraphics[angle=0,scale=0.3]{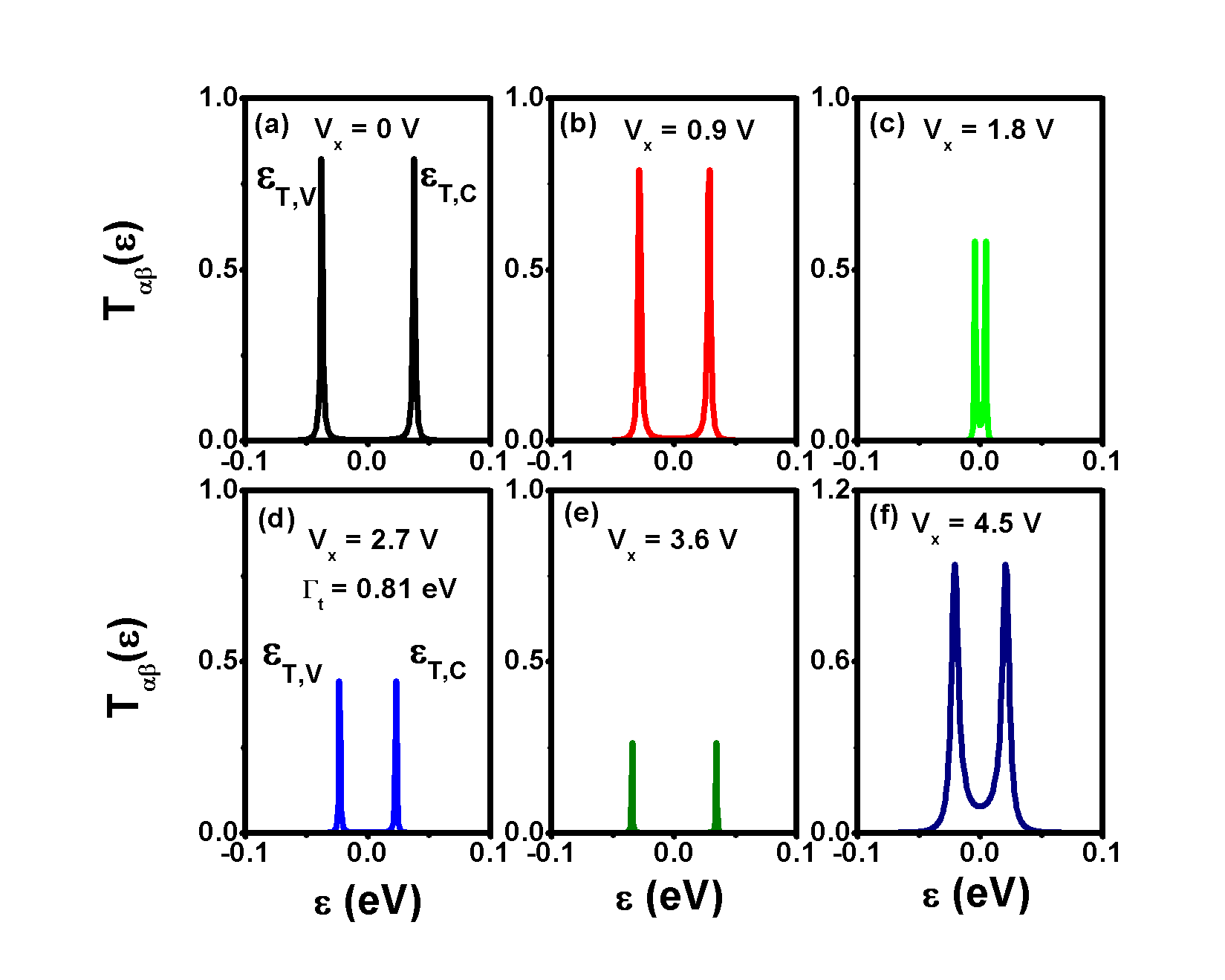}
\caption{Transmission coefficient ${\cal
T}_{\alpha,\beta}(\varepsilon)$ of $9_8-7_5-9_8$ AGNR segments for
various electric fields at $\Gamma_t = 0.81$ eV. Diagrams (a)-(f)
correspond to $V_x = 0$ V, $V_x = 0.9$ V, $V_x = 1.8$ V, $V_x =
2.7 $ V, $V_x = 3.6 $ V, and $V_x = 4.5 $ V.}
\end{figure}

\section{Conclusion}
We have conducted a theoretical investigation into the Stark
effect on the topological states (TSs) of finite AGNRs and
heterostructures. These TSs encompass the multiple end zigzag edge
states of 13-AGNRs and the interface states of $9_w-7_x-9_y$ AGNR
heterostructures. The multiple end zigzag edge states exhibit a
blue Stark shift in their energy levels. Analysis of the charge
density distribution reveals that the electric field aligns with
the dipole moment of their electronic structures. In $9_w-7_x-9_y$
AGNR segments, both blue and red Stark shifts can occur.
Specifically, in $9_4-7_3-9_4$ AGNR segments, the blue Stark shift
is observed under small electric fields, while in $9_8-7_3-9_8$
AGNR segments, the red Stark shift is observed. However, these
Stark shifts exhibit a common phenomenon, namely an oscillatory
behavior around the Fermi energy concerning transverse electric
fields.

Furthermore, we have computed the transmission coefficients
(${\cal T}_{\alpha,\beta}(\varepsilon)$) of these two types of TSs
under transverse electric fields. The spectra of ${\cal
T}_{\alpha,\beta}(\varepsilon)$ for 13-AGNRs reveal peak
magnitudes within the band gap larger than one for various
transverse electric fields. This demonstration indicates that
finite 13-AGNRs featuring multiple end zigzag edge states can be
observed when armchair edge sites are coupled to the electrodes.
Moreover, the ${\cal T}_{\alpha,\beta}(\varepsilon)$ of
$9_8-7_5-9_8$ AGNR segments illustrates that the topological
interface states between the 9-AGNR and the 7-AGNR segments can be
effectively modulated by transverse electric fields. Such
characteristics hold significant promise for employing these TSs
{\color{red} as charge
qubits[\onlinecite{ChenCC},\onlinecite{Trauzettel}]}.


{}

\textbf{Data availability statement} All data that support the
findings of this study are included within the article(and any
supplementary files.)

{\bf Acknowledgments}\\
This work was supported by the National Science and Technology
Council, Taiwan under Contract No. MOST 107-2112-M-008-023MY2.

\mbox{}\\
E-mail address: mtkuo@ee.ncu.edu.tw\\

 \numberwithin{figure}{section} \numberwithin{equation}{section}

\setcounter{equation}{0} 

\mbox{}\\

\numberwithin{figure}{section}



\end{document}